# Neck Barrier Engineering in Quantum Dot Dimer Molecules via Intra-Particle Ripening

Jiabin Cui[1,2+§], Somnath Koley[1,2§], Yossef E. Panfil[1,2], Adar Levi[1,2], Yonatan Ossia[1,2] Nir Waiskopf[1,2], Sergei Remennik[2], Meirav Oded[1,2] & Uri Banin[1,2]*

1 Institute of Chemistry, The Hebrew University of Jerusalem, Jerusalem 91904, Israel.

2 The Center for Nanoscience and Nanotechnology, The Hebrew University of Jerusalem, Jerusalem 91904, Israel.

§J.C., and S.K contributed equally to this work.

**Abstract**: Coupled colloidal quantum dot (CQD) dimers represent a new class of artificial molecules composed of fused core/shell semiconductor nanocrystals. The electronic coupling and wavefunction hybridization is enabled by the formation of an epitaxial connection with a coherent lattice between the shells of the two neighboring quantum dots where the shell material and its dimensions dictate the quantum barrier characteristics for the charge carriers. Herein we introduce a colloidal approach to control the neck formation at the interface between the two CQDs in such artificial molecular constructs. This allows the tailoring of the neck barrier in pre-linked homodimers formed via fusion of multifaceted wurtzite CdSe/CdS CQDs. The effects of reaction time, temperature and excess ligands is studied. The neck filling process follows an intraparticle ripening mechanism at relatively mild reaction conditions while avoiding inter-particle ripening. The degree of surface ligand passivation plays a key role in activating the surface atom diffusion to the neck region. The degree of neck filling strongly depends also on the initial relative orientation of the two CQDs, where homonymous plane attachment allows for facile neck growth, unlike the case of heteronymous plane attachment. Upon neck-filling, the observed red-shift of the absorption and fluorescence measured both for ensemble and single dimers, is assigned to enhanced hybridization of the confined wavefunction in CQD dimer molecules, as supported by quantum calculations. The fine tuning of the particle interface introduced herein provides therefore a powerful tool to further control the extent of hybridization and coupling in CQD molecules.

**INTRODUCTION**

Colloidal Quantum Dots (CQDs), often referred to as "artificial atoms", have reached a high level of materials control, yielding a comprehensive library of high-quality nanocrystals with varying size, shape, and composition. [1-4] Promoting the concept of "nanocrystals chemistry" in which the CQDs acting as artificial atoms, are assembled and fused to form artificial molecules [5-7] and artificial solids [8-11] is a promising route to new nanocrystals based materials with novel functionalities. The organic ligands coordinated at the CQD surface act as a potential barrier that minimizes the delocalization of the charge carrier wavefunctions and hence impedes coupling between neighboring CQDs in such constructs. Epitaxial fusion between neighboring CQD facets was addressed especially in CQD solids. The aim is to form a continuous coherent crystal that lowers the energy barrier between two adjacent CQDs via the necking process, thus leading to stronger coupling due to enhanced overlap of the carrier wavefunctions.[5,9-12] The width of the neck is an impactful parameter in governing the carrier coupling across the CQD solids and is generally dictated by the nature of the crystal facets in the monomer nanocrystals and the attachment scenario. Solvent evaporation control and colloidal atomic layer deposition (C-ALD), were utilized towards controlling necking between the adjacent CQDs especially for softer semiconductor lattices.[13-18] These "connected but confined" superstructures show promise for CQD based field effect transistors, sensors, photodetectors, and future generation quantum devices. [15-21]

Recently we introduced coupled colloidal quantum dot molecules constructed from prototypical CdSe/CdS core/shell CQDs forming dimers that were fused at moderate temperatures. Using small cores with thin shells allowed for extensive electronic coupling in the fused dimers showing a plethora of coupling signatures, most significantly red shifting and broadening of the absorption and fluorescence.[5] While the center to center distance between the cores is an important parameter governing the extent of exciton wavefunction overlap in such dimers, the neck characteristics are also central for tuning the extent of coupling and wavefunction hybridization.[5-6, 22] Beyond the monomer shell thickness, which in the fused dimers determines the barrier thickness, the barrier width creating a constriction between the two CQDs, is an additional critical parameter dictating the coupling in CQD molecules. Neck parameter as mentioned before is also crucial for the long range electronic coupling, studied mostly for the 'soft' semiconductor systems. In order to precisely tailor the interfacial barrier in fused CQDs, several important questions need to be addressed. Can the neck be filled beyond the facet size? If so, to what extent do we change the inherent properties of the monomer? What are the governing factors towards a general route? Hence a systematic and comprehensive study to tailor the neck barrier at the CQD interface is warranted to address these questions and achieve the highest possible electronic coupling. A novel mechanism with well-understood and general working principle can potentially address the barrier control in coupled CQD systems enriching the nanoscale design parameters.

Herein, starting from chemically cross-linked CdSe/CdS core/shell homodimers, we achieve control over the fusion step while examining the effects of the reaction temperature, time and the role of excess ligands on the resultant neck width. The interplay between the reaction temperature and surface ligand density yields dimer architectures with different neck characteristics, ranging from weakly-fused to a rod-like structure, while under similar reaction conditions the monomers are not significantly altered.

Our approach encompasses fine tuning of intraparticle ripening via diffusion of surface atoms within the nanocrystal dimer while attempting to suppress undesired inter-particle ripening which could compromise the dimer structures. Intraparticle ripening is governed by the relative reactivity of the specific facets and sites in the architecture [23]. It has been utilized in shape control of semiconductor nanocrystals to transition from spherical to tetrahedral CQD architectures [24], and in controlled metal tip growth on nanorods.[25-27] Notably, various inter-particle ripening mechanisms between different nanocrystals are also common. Ostwald ripening is the most prominent, where larger nanocrystals are formed at the cost of dissolution of smaller nanocrystals, typically occurring at low ligands concentration and higher temperature [27-28]. Additionally, digestive ripening occurs at a ligand rich environment, where the added free ligands accelerate the ion solubility at the nanocrystal surface and beyond a moderate thermal and particle concentration threshold this leads to increase in particle size and reshaping.[29] Both of these mechanisms ought to be avoided for intraparticle rearrangement within a dimer so as not to significantly compromise the inherent size, shape and composition of the original monomers.[30-32]

Indeed, a suitable working window had to be found to enable systematic tuning of the neck barrier without any significant alterations of the core-location, center-to-center distance and shell thickness in the CQD dimer molecules. Correspondingly, this type of control over the neck characteristics, enabled tuning of the electronic coupling and hence the hybridization of the electron wavefunction in the dimers, allowing for efficient inter-dot coupling reproduced also via quantum mechanical calculations.

**Results and Discussion**

**Neck filling strategy in CQD dimer composed of model CdSe/CdS CQDs**

CdSe/CdS core shell nanocrystal CQDs, represent a model system, with quasi-type-II band alignment where the electron wavefunction tends to delocalize throughout the core/shell structure due to its small effective mass and low core-shell conduction band offset, while the heavier hole experiencing larger valence band offset, is confined to the core (Figure 1a).[33] We focused on CQD dimers that can show quantum-coupling and wavefunction hybridization, constructed of core/shell monomers with 2.8 nm diameter CdSe core and moderate CdS shell thickness (6ML) with an overall diameter of ~6.8nm (Figure S1, Figure S2) as identified by the transmission electron microscopy (TEM) and high-angle annular dark field-scanning TEM (HAADF-STEM). These CQDs fall well in the quasi type-II regime, as indicated by the redshift in the band-edge absorption and fluorescence upon shell growth, accompanied by an elongated exciton lifetime.

Fusing two such CQDs together, leads to the hybridization of the electron wavefunction in the artificial molecules that are formed.[5-6, 34-36] The synthesis of the artificial molecules follows in general the procedure described by us recently, while the fusion step was systematically modified to control the extent of neck filling in the attained dimers.[5] Briefly, silica spheres with diameter of ~200nm were used as a template for the controlled dimers formation. The first monomer layer is bound to the thiol functionalized silica sphere surface, followed by masking of the surface with a thin silica layer to fixate the monomers and block further binding to the silica itself. The bound CQDs are functionalized by tetra-thiol molecules serving as a linker. Then the solution is exposed to the second CQD layer, which selectively bind to the first CQDs through the thiol linkers. In the next step, the silica spheres are etched selectively by HF, and the solution is cleaned from silica fragments. Figure S3 shows a purified fraction of non-fused dimers after etching. The solution is next taken to the fusion step discussed below. The barrier of the

In the critical fusion step, in order to control the neck dimensions, we carried out a study of the effects of the fusion reaction conditions in octadecene (ODE) as illustrated in Figure 1b, in the presence of a fixed amount of Cd-oleate, added in order to stabilize the dimers while suppressing the interparticle ripening. This included varying the reaction temperature between 120-240 °C, along with the judicious addition of coordinating ligands (oleylamine (OAm), oleic acid (OA)) to assist in the dimer surface passivation and its solution dispersion (Table S1, includes only the conditions resulting in substantial neck filling, at reaction temperature above 200 °C). After the thermal fusion step, the solution undergoes size-selective precipitation using solvent/antisolvent combination, to attain the purified dimers fraction while removing remaining monomers and higher order oligomers. The replacement of the organic ligands and the environmental barrier with CdS nanocrystal lattice expectedly facilitate the gradual enhancement in delocalization of the electron wavefunction proportionately in the CQD molecules (Figure 1c).

Starting from organically cross-linked dimers, promoting the facet attachment via the thermo-chemical fusion strategy requires the diffusion of surface atoms. This process was found to require temperature above 180 °C in presence of the fixed amount of Cd-oleate. Indeed, as seen in Figure 1d, e, and f, at a temperature of 120 °C, no fusion is observed. Only at 200 °C, with ~150μM of passivating ligand (OAm+OA) concentration (Fig 1d, e, f), a reasonable attachment of the facets was achieved. Monomer particles under similar reaction conditions did no show any change. Therefore, we conclude that the neck fusion and filling process is dominated by intraparticle ripening involving the atoms on the dimer itself.

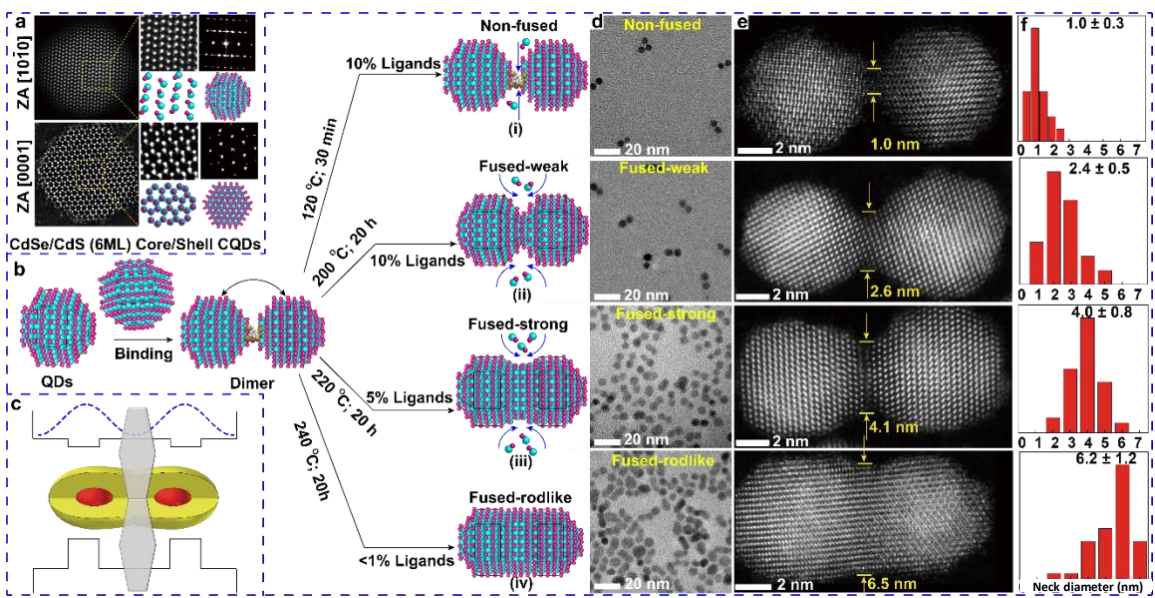

**Figure 1. Progressive neck filling in CQD dimer molecules.** (a) Atomic level detailed characterization of monomer CdSe/CdS CQDs. The nanocrystals possess phase pure wurtzite structure with no interfacial defects or stacking faults. (b) The synthetic scheme for the neck barrier control in CQD dimers varying the fusion reaction conditions. Interplay between the reaction temperature and added coordinating ligands was studied as the key variables. (c) Cartoon representation illustrating the effect of neck barrier on hybridization of electron wavefunctions among neighboring particles. (d) TEM and (e) HAADF-STEM images for dimer particles fused under the different conditions. The neck diameter was determined from HAADF-STEM images and the statistical data is shown in (f).

However, the neck diameter following the aforementioned reaction condition is apparently limited, thus resulting in a neck diameter of ~3nm (Figure S4-S5), a so-called "weak fusion" situation, where the neck diameter is generally limited by the fused facet size. Further filling of the neck region was accomplished by tuning of the fusion reaction conditions. At elevated temperature (~220-240 °C) with the ligand concentration kept the same (~150μM OAm+OA), a similar weak fusion result was observed. Beyond 240 °C, significant inter-particle ripening was observed, compromising the dimer structure. Within the relevant fusion reaction temperature range, we found that at 220-230 °C, while lowering the surface ligand concentration to tens of μM, leads to a larger neck diameter of 4-4.5nm, so-called "strong fusion" with peanut-like dimer architecture. We assign this to the reduced degree of surface passivation, facilitating the surface atom movement to the neck region increasing its diameter beyond the facet size (Figures 1d, e; S6-S7). Considering that a deficiency of the surface ligands in tandem with elevated reaction temperature promotes the intra-particle ripening, we raised the temperature further to 240 °C while keeping the free ligand concentration low (at ~7 Mm). Indeed, this completes the filling of the neck, yielding a rod-like dimer structure with coherent lattice throughout (Figure 1d, e; S8). Further increase in temperature to above 240 °C in the ligand deprived conditions, promoted the Ostwald inter-particle ripening extensively for both the monomer and dimers (Figure S9).

In all the aforementioned fusion reaction conditions, and most significantly even at the rod-like fusion conditions, EDS-STEM characterization of the dimers, shows that the continuous shell embraces the two cores at a similar distance as in dimers fused in the other conditions (Figure 2a-c, S4, S7). The elemental mapping indicated the preservation of the two CdSe cores, covered by CdS shell whose thickness is maintained, confirming that the intraparticle ripening approach affected mainly the neck diameter allowing its control from weakly-fused to the rod-like dimer CQDs.

Recapping, the intermediate reaction conditions described above facilitate adequate control on neck diameter, thus providing a knob to tailor the extent of hybridization in CQD molecules. The observed behavior suggests that at the thermo-chemical fusion condition, the surface atoms rearrange by moving to the dimer interfacial neck region, where there is surface strain due to interfacial curvature that increases the surface energy and thus promotes the formation of a more continuous structure.

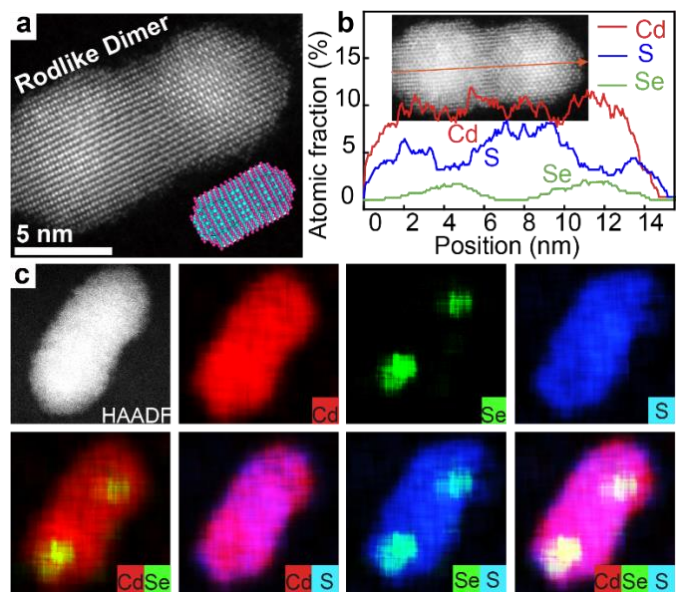

**Figure 2. Structural aspects of the neck filled dimers.** (a) HAADF-STEM image, (b) EDS line scan, and STEM-EDS mapping for elemental analysis of core and shell properties in the neck filled rod-like dimer. Core location and shell properties show the core-to-core distance remain similar as in weakly fused dimer, without additional strain altering band-edge properties.

**Structural Aspects of the Neck-Filled Dimer CQDs:**

Going deeper into the consequences of the attachment scenario followed by the fusion reactions, it is important to consider the hexagonal wurtzite morphology of the monomer CQD units manifesting multiple surface facets (namely, (002), (100), (101), Figure S1) with different surface energies dictated by the number of dangling bonds. This plays a role in the cross-linking step, and the relative affinity of the surface Cd atoms to the thiol head group of the tetra-thiol cross-linker, leads to a distribution of dimers with either homonymous (same facet linking and fusion) or heteronymous plane attachment (linking and fusion of different facets).[14] [30]

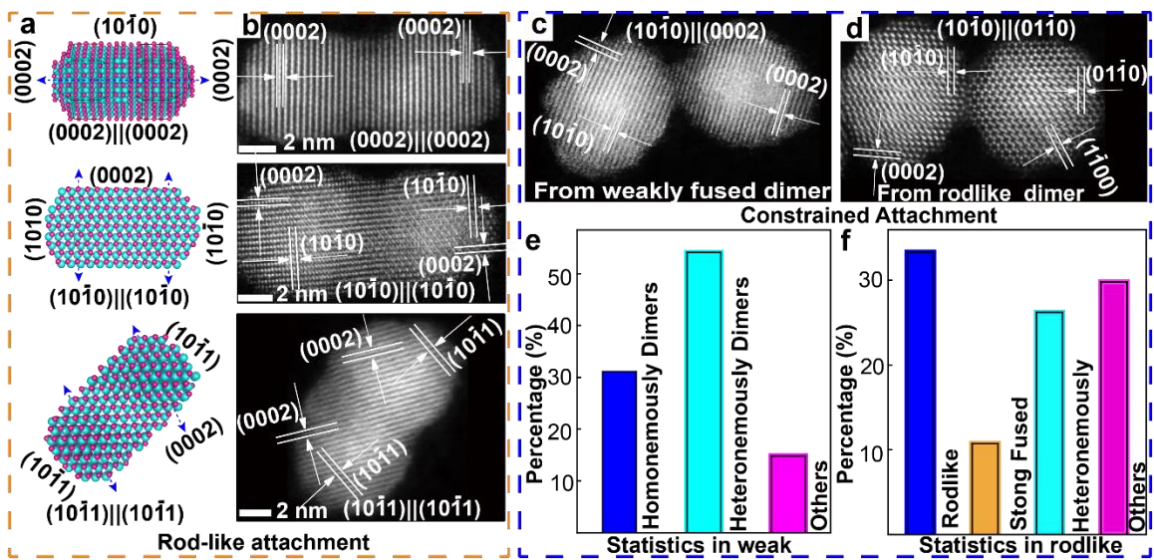

**Figure 3. Effect of various attachment motifs on the neck filling of CQD dimer molecules.** 3D reconstructed models (a) and corresponding HRTEM images (b) for homonymous attachment via (002), (100), and (101) crystal planes leading to similar neck filled dimer following robust fusion conditions. Heteronemously attached dimers following (c) mild and (d) robust fusion conditions. In this case, the interface is strained following dissimilar facet attachment restricting facile atom rearrangement at the neck region. Statistical occurrences of various species following (e) mild, and (f) robust fusion conditions.

Above, we described the successful filling of the neck via optimized reaction conditions focusing on the homonymously attached dimer CQDs. Figure 3a-b represents three abundant types of rod-like dimers connected via ($10\bar{1}1\|10\bar{1}1$), ($10\bar{1}0\|10\bar{1}0$), and ($0002\|0002$) facets respectively, that are also three major homonymous plane attached population in case of weakly fused dimer. Following the slow reaction pathway, neck filling in the homonymous attachment case is completely irrespective of the crystal facet reactivity. Nonetheless, a more perfect rod-like structure was observed with ($10\bar{1}0\|10\bar{1}0$) attachment alignment. The neighboring facet of the neck area in this case is (0002) which possess a larger number of dangling bonds (Figure S2) and hence the movement of the surface atoms during intraparticle ripening easily facilitates neck filling to achieve a rod-like geometry.[30]

Figure 3c-d presents cases of heteronomous attachment, where the crystal planes are not aligned, owing to the strained particle-to-particle interface (Figure 3c-d; S10). In such cases, the neck properties of the constrained dimer do not change significantly even at the rod-like fusion reaction conditions.

The statistical distribution of the different species is presented in Figures 3e, f, for the weakly fused and rod-like dimers. The population of homonymously attached dimer are converted to rod-like dimer with similar abundance of over 30% (blue bars in Figure 3e, f). However, only a certain population of the heteronymously attached dimer are fused in a better way leading to a strongly fused dimer with an intermediate neck (yellow bar in Figure 3f). Healing of the defected interface at a high temperature can cause this type of rearrangement.[18] Other structures, such as trimer or higher order oligomers are also more probable in the rod-like fusion conditions, which are later separated by size-selective precipitation.

**Spectral Feature and Electronic Coupling in Neck Filled CQD Dimer Molecules**

We next consider the neck effects on the quantum coupling and wavefunction hybridization in the artificial molecules. The coupling is experimentally evidenced by the red-shift in the band edge properties providing a measure of the delocalization and splitting of the electron wavefunction into bonding and antibonding states.[5-6]

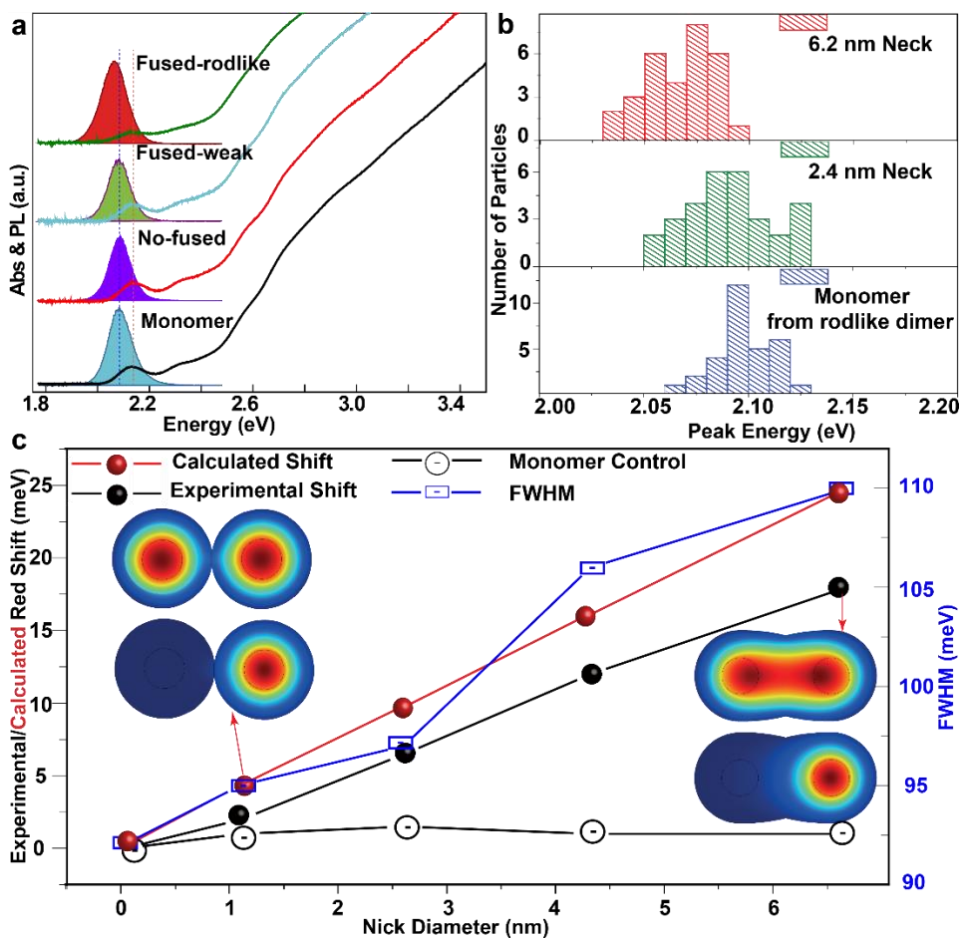

**Figure 4. Optical and electronic properties of CQD dimer molecules with neck barrier variation.** (a) Absorption and emission spectra of various neck filled dimers with respect to the monomers. A gradual red-shift and broadening of both the band-edge emission, along with enhanced absorption at the CdS shell region evidence the formation of a continuous coherent crystal. (Mid-neck dimers' spectra are not shown for clarity). (b) Single particle emission peak position as studied with processed monomers, weakly fused dimer, and rod-like dimers. (c) Wavefunctions and band-edge red shift as a function of neck diameter; Filled circles represents experimental (black) and calculated (red) shift as a function of neck diameter. A control experiment was performed with the monomer spectra undergoing similar reaction conditions (void circles). Blue squares represents the spectral width of different neck-filled dimer CQDs.

Figure 4a presents the ensemble absorption and emission spectra for the monomers and dimers produced via different fusion conditions. A red shift and broadening in the absorption and emission are observed after fusion, becoming more prominent with increased neck size. In the rod-like neck regime, a significant increase in CdS absorption peak was also observed along with the alterations at the band-edge. This can be attributed to the development of a continuous CdS shell, increasing the oscillator strength of the shell related particular optical transitions. Since the organic linker barrier is replaced by a continuous crystal, the delocalization of the electron wavefunction into the neighboring CQDs become facile which increases the electronic coupling strength. Single particle emission spectra were also recorded for the different dimer samples as shown in Figure 4b (see SI for further details). A systematic red shift of the center wavelength in the distribution upon filling of the neck is seen.

Quantum-mechanical calculations were performed to validate the effects of the neck filling on the quantum coupling and wavefunction hybridization. We have calculated the energy levels and the wave-functions, as a function of the neck barrier, using single band effective mass approximation and the self-consistent Schrodinger-Poisson equations (see SI for further details). This allowed to extract the red shift in the band gap relative to the non-fused dimers, while considering the energy difference between the symmetric and anti-symmetric states including the electron-hole Coulomb interaction. The wavefunctions for two different neck diameters (1 nm and 4.2 nm) are presented in the inset of Figure 4b. The electron wavefunction delocalization is significantly enhanced in the latter case, leading to a larger red-shift. As the neck filling proceeds, the delocalization of the electron wavefunction also affects the electron-hole coulomb interaction, which reduces the observed shift compared to that expected for the effect of hybridization alone. A linear response of the observed shift as a function of the neck diameter is calculated, in good agreement with the experimental trend.

The results of the experimental ensemble and single particle red shifts are compared with the calculated values. As discussed before, the separated monomer that underwent similar reaction conditions from each set of dimers, do not exhibit any considerable red-shift in the band-edge properties (Figure S11). This was also verified by examining the single particle emission characteristics of the separated monomers from different dimer sets. The monomers that underwent the rod-like fusion condition, exhibit classical ON-OFF blinking with high ON fraction, and characteristic single exponential radiative lifetime (Figure S12). A reasonable agreement is seen between experimental and calculated shifts. While the ensemble spectra shows a slightly smaller red-shift at larger neck diameter (17 meV observed compared to calculated 24meV), the single particle spectra is in better agreement with the calculated red-shift, where ~25-30meV shift is observed in rod-like dimer (Figure 4b). A presence of dimers with inadequate neck filling in the population can contribute to this effect in the ensemble measurements. The spectral width of the emission of the dimers is significantly broadened above a neck width of 4 nm (Figure 4c), while for the separated monomers the width is unaltered. Two factors contribute to this, (1) the delocalization of the electron wavefunction leads to weaker band edge confinement leading to broader spectra as a manifestation of coupling; (2) Presence of two populations in the neck filled dimer sample, where the heteronymously attached dimer remain weakly fused, exhibiting smaller red-shift and thus broadening the overall dimer spectral feature.

**Mechanistic Insights and Summary**

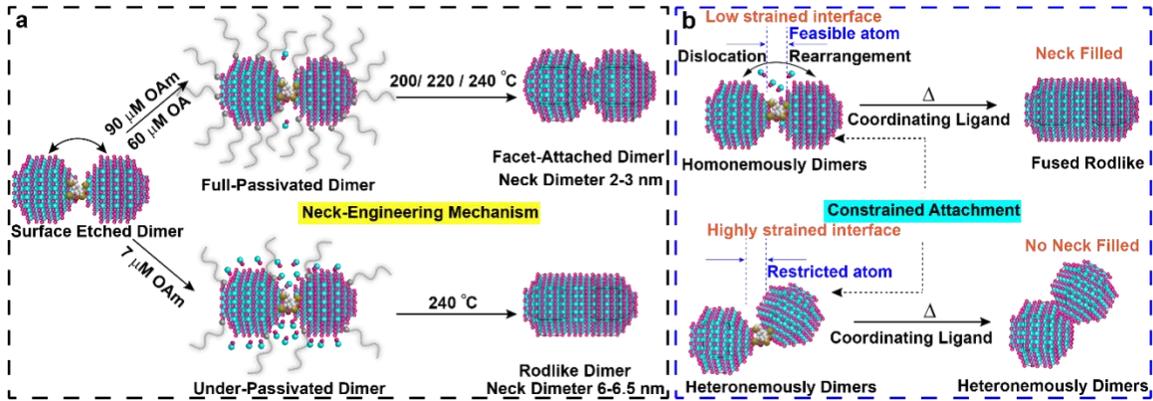

**Figure 5. Neck-Engineering mechanism.** (a) Role of surface passivation on the intra-particle ripening mechanism. Starting from a surface etched cross-linked dimer, a full passivation with coordinating ligands reduces the surface atom diffusion, and hence a limited neck diameter is observed at different temperature ranges. While in ligand deprived regime, the surface atom diffusion is activated and a full neck filling at high temperature threshold can be achieved. (b) Effect of constraints during initial attachment; Neck filling for dimer with continuous and strained interface.

The intra-particle ripening mechanism for neck filling of the CQD dimer relies largely on the surface chemistry. In the case of a fully ligand passivated dimer, the motion of the surface atoms is limited in the intra-particle regime. In that case, the fusion reaction leads to only facet attachment with no neck filling independent of the temperature at the range of 200-240 °C (Figure 5a). The effect of temperature becomes more obvious when the surface of the dimer is not fully passivated. The under-passivated atoms can more easily disconnect and migrate towards the neck region while the thermal threshold remains lower than the inter-particle ripening regime. Under this regime of poor surface passivation, the neck diameter shows a proportionate behavior with reaction temperature and time (Table S1).

The template based synthesis of nanocrystal dimers leads to both homonymous and heteronymous attachment of the nanocrystal facets, depending on the facet-thiol reactivity (Figure 5b). Once the two nanocrystals are crosslinked, they cannot further rotate to align different facet, unlike the oriented attachment of the free CQDs in solution.[37-39] We have discussed earlier that the intraparticle ripening can feasibly occur in the homonymously attached CQD dimers irrespective of the plane of attachment. In this case, a stable interface is formed within the two CQDs, where the surface energy remains higher, leading to movement of the atomic precursors to fill the neck region. In the mild fusion condition, i.e. in the weakly fused dimers' set, a certain population of heteronymously attached dimer was observed, although those possess similar neck size compared to the homonymously attached dimers. In the stronger fusion condition, predominantly rodlike architecture was obtained for the homonymously attached dimer. However, certain population exhibited to possess weak neck (~3nm). When examined via high resolution STEM, we found that almost all of them are heteronymously attached. In case of the constrained attachment scenario, the relative facets orientation is governed by the binding of the linker molecules at first to the exposed monomer facet on the silica template, and then to the secondary binding of the added monomers in the dimer formation step. In the case of heteronymous orientation (e.g. Fig. S10), the atoms at the dimer interface are strained, and the strain energy restricts further rearrangements at the neck region, and hence the neck size remains limited to the attached common area of the two nanocrystals. Nonetheless, such treatment can anneal the interface to a smoother one, removing imperfections in some cases.[18]

Notably, the aforementioned intra-particle ripening mechanism works well for the homonymous oriented dimers.

**Conclusion:**

We find that the extent of electronic coupling in CQD molecules can be tuned systematically, by a synthetic strategy to control the fusion reaction. Although initially attached CQDs, form a limited neck governed by the size of the facets, a non-digestive intraparticle ripening condition allows the movement of the surface atoms towards the dimer interface region. We have set the region of the feasible neck filling strategy without activating the inter-particle ripening mechanism, leading to stronger electronic coupling among two CQD artificial atoms. This knowledge can potentially create the pathways to enhance long range electron delocalization in connected-but-confined artificial atom superstructures, a potential candidate for quantum information science.[40]

**AUTHOR INFORMATION**

Corresponding Author

* Email: uri.banin@mail.huji.ac.il (U.B.).

Authors

[+]Present address: The Center for Molecular Imaging and Nuclear Medicine, State Key Laboratory of Radiation Medicine and Protection, School for Radiological and Interdisciplinary Sciences (RAD-X) and Collaborative Innovation Center of Radiological Medicine of Jiangsu Higher Education Institutions, Soochow University, Suzhou 215123, China.

Author Contributions

[§]J.C., and S.K contributed equally to this work.

Notes
The authors declare no competing financial interest.

**ACKNOWLEDGMENT**

The research leading to these results has received financial support from the European Research Council (ERC) under the European Union's Horizon 2020 research and innovation programme (grant agreement No [741767]). J.C. and S.K acknowledge the support from the Planning and Budgeting Committee of the higher board of education in Israel through a fellowship. U.B. thanks the Alfred & Erica Larisch memorial chair. Y.E.P. acknowledges support by the Ministry of Science and Technology & the National Foundation for Applied and Engineering Sciences, Israel.

# Supplementary Information

**Materials and Methods:**

*Reagents:* Oleylamine Oleylamine (OAm, 70%), trioctylphosphine oxide (TOPO, 99%), 1-octadecene (ODE, 90%), oleic acid (OA, 99%), 1-octanethiol (98.5%), selenium (Se, 99.99%), cadmium oxide (CdO, 99.99%), pentaerythritol tetrakis(3-mercapto-propionate) (PTMP, 95%), ammonia aqueous (28.5%), N-methylformamide (NMF, 99%), tetrahydrofuran (THF, 99.9%), toluene (99.8%), polyvinylpyrrolidone (PVP, 10k), hydrofluoric acid (HF, 48%), ethanol (99%), hexane (95%) (3-Mercaptopropyl) trimethoxysilane (MPTMS, 95%) and tetraethyl orthosilicate (TEOS, 98%) were obtained from Sigma Aldrich. Trioctylphosphine (TOP, 97%) was purchased from Strem Chemicals. Octadecylphosphonic acid (ODPA, 99% was purchased from PCI synthesis. All the reagents were used as received without further purification.

*CdSe core growth:* The wurtzite CdSe core was synthesized according to our previous method with a slight modification[1,2]. Briefly, 6 g TOPO, 560 mg ODPA, and 120 mg CdO were added to a 50 mL flask and degassed at 120 °C for 30min. Then the mixture was heated to 350 °C under Ar flow. When the color of the solvent changed from brown to colorless, 1 mL of TOP and 1 mL of Se/TOP (120 mg of Se in 1mL TOP) was swiftly injected into the flask, respectively. After 30 s, the heat-mantle was directly removed and cool down to room temperature with blower. Then the resulting CdSe nanoparticles (NPs) were washed through centrifugation for 3 time and redispersed in 10 mL hexane for the next shell growth step.

*The synthesis of wurtzite CdSe/CdS core-shell colloidal quantum dots:* The concentration and core size of CdSe was identified by the absorption spectrum[3]. Then 100 nmol of CdSe colloidal quantum dots (CQDs) was added to a 50 mL flask and mixed with 3 mL OAm and 3 mL ODE. The reaction solution was degassed at 120 °C for 1h to remove the hexane and oxygen. Then the reaction solution was heated up to 310 °C under argon flow. During this time, 6 mL of Cd-oleate and octanetiol precursor was injected dropwise into the solution at 240 °C with a rate of 3 mL/h[4]. After the shell growth reaction, 1 mL of OA was injected into the flask to passivate the CdSe/CdS core/shell QDs for 30 min. Finally, the wurtzite CdSe/CdS QDs were collected by centrifugation and keep in 10 mL hexane.

*SH-SiO$_2$ NPs synthesis:* The SH-SiO$_2$ NPs were synthesized with two step. Firstly, 180 mL of ethanol and 17 mL of NH$_3$•H$_2$O (28%) were added in the flask and stirring with the rate of 1200 rpm/min for 5min[5]. Then 5 mL of TEOS was added into the reaction solution. After stirring for 10 h, 1.5 mL of MPTMS was injected and keep stirring for another 10 h. Finally, the SH-SiO$_2$ NPs were collected by centrifugation and redispersed in 31 mL of ethanol.

*The formation of dimer CdSe/CdS@SiO$_2$*: The CdSe/CdS QDs dimer structure was synthesized according to our previous method[1,2]. Firstly, 1 mL of SH-SiO$_2$ NPs was mixed with 0.1 mL PVP/ethanol solution (20 mg/mL) for 1h and redispersed in 3 mL of toluene. During this time, 2 nmol of wurtzite CdSe/CdS QDs was mixed with the cleaned SiO$_2$-PVP solution for 1 h, yielding the CdSe/CdS@SiO$_2$. After washing for 2 times, CdSe/CdS@SiO$_2$ was dispersed in 5 mL of ethanol. Then 330 μL of ammonia solution (28.5% wt %) was added into the solution with stirring for 5 min. Following, 20 μL of TEOS was injected dropwise by the syringe pump in 2 h. After stirring for another 3 h, the samples were collected by centrifugation and redispersed in 5 mL of THF, forming the SiO$_2$@CdSe/CdS@SiO$_2$ with surficial silica layer for the masking and immobilization. Subsequently, 100 μL of PTMP was added into the above solution with stirring for 10 h. After centrifugation for 2 times, the samples were dispersed in 5mL of THF. Next, 2 nmol of CdSe/CdS QDs was added dropwise and keep stirring for 3 h. Finally, the dimer CdSe/CdS@SiO$_2$ were collected by centrifugation procedure.

*The etching and release of CdSe/CdS QDs Dimer:* The CdSe/CdS QDs dimer was fabricated by the etching and release method. Briefly, dimer CdSe/CdS@SiO$_2$ was mixed with 2 mL of HF/NMF (15%) solvent under stirring in the oil bath at 60 °C for 10 h. Then the CdSe/CdS QDs dimers were collected by centrifugation and dispersed in 2 mL of ethanol.

*The synthesis of fused CdSe/CdS CQDs structure:* The conditions for the fused CdSe/CdS QDs with different morphology are shown in the Table S1. Taking the rod structure for example, CdSe/CdS dimer (in 2 mL of ethanol) were mixed with 3 mL of ODE, 300 μL of Cd-oleate (0.2 M), and 5 μL of OAm. The reaction solution was degassed under vacuum at room temperature for 10 min and again at 120 °C for an additional 30 min. Later, the reaction mixture was heated to 240 °C for 20 h under argon flow. The resulting fused particles were precipitated with ethanol and dispersed in 2 mL toluene.

*Characterization:* Absorption spectra were measured using a Jasco V-570 UV-Vis-NIR spectrophotometer. Fluorescence spectra was measured with a fluorescence spectrophotometer (Edinburgh instruments, FL920). Transmission electron microscopy (TEM) was performed using a Tecnai G$^2$ Spirit Twin T12 microscope (Thermo Fisher Scientific) operated at 120 kV. High-resolution TEM (HRTEM) measurements were done using Tecnai F20 G$^2$ microscope (Thermo

Fisher Scientific) with an accelerating voltage of 200 kV. High-resolution scanning-transmission electron microscopy (STEM) imaging and elemental mapping was done with Themis Z aberration-corrected STEM (Thermo Fisher Scientific) operated at 300 kV and equipped with high angular annular dark field detector (HAADF) for STEM and Super-X energy dispersive X-Ray spectroscopy (EDS) detector for high collection efficiency elemental analysis. The images and EDS maps were obtained and analyzed with Velox software (Thermo Fisher Scientific).

*Single Particle Measurement:* Single Particle measurements was performed with an inverted fluorescence microscope in epi-luminescence configuration. The diluted particles were spin coated on clean glass coverslips. 405nm pulsed laser was focused on the particle with oil-immersion objective (100X; 1.4NA). The fluorescence from the particle was collected with the same objective and passed through a dichroic mirror (550nm), and a long-pass filter (570nm) to eliminate the excitation light. The emission light was focused to a fast EMCCD camera (Andor iXon3) through a spectrograph (Acton-Pi). The spectrum was recorded at 80mS/frame rate and integrated over 120s. The peak positions are reported in the main text (Figure 4c). To identify the particles as monomer or dimer, we checked the same particle by sending the emission to Avalanche Photodiode, measuring their time-tagged-time-resolved data. As reported in our earlier paper[2], monomer exhibits bi-modal intensity distribution, long lifetime of the ON state, while dimer possess relatively higher emission intensity, flickering of emission, shorter lifetime.[1] We have measured the single particle emission spectra from the weakly coupled and strongly coupled dimer sets, where we could identify the particles from sets of the variable electronic coupling. The other photophysical singnatures from the weakly coupled dimer sets are noted while measuring the single paricle spectrums. The presence of weakly coupled dimer in the rodlike sample set thus was identified to record the emission spectrum from the strongly coupled particles only. The monomer data shown here are the monomers separated from the dimer sample set, to consider the exact shift due to experimental conditions.

Table S1 Details of fusion parameters for different dimer sets.

| Reaction Temp (°C) | Fusion Time (h) | [OAm] (μM) | [OA] (μM) | Neck Diameter (nm) |
|---|---|---|---|---|
| **200** | 20 | 90 | 60 | 2.5-3 |
| **220** | 20 | 90 | 60 | 2.5-3 |
| **240** | 20 | 90 | 60 | 2.5-3 |
| **220** | 20 | 45 | - | 4-4.5 |
| **220** | 20 | 7 | - | 4-4.5 |
| **230** | 20 | 7 | - | 4-4.5 |
| **240** | 20 | 7 | - | 6-6.5 |
| **240** | 30 | 7 | - | 6-6.5 |

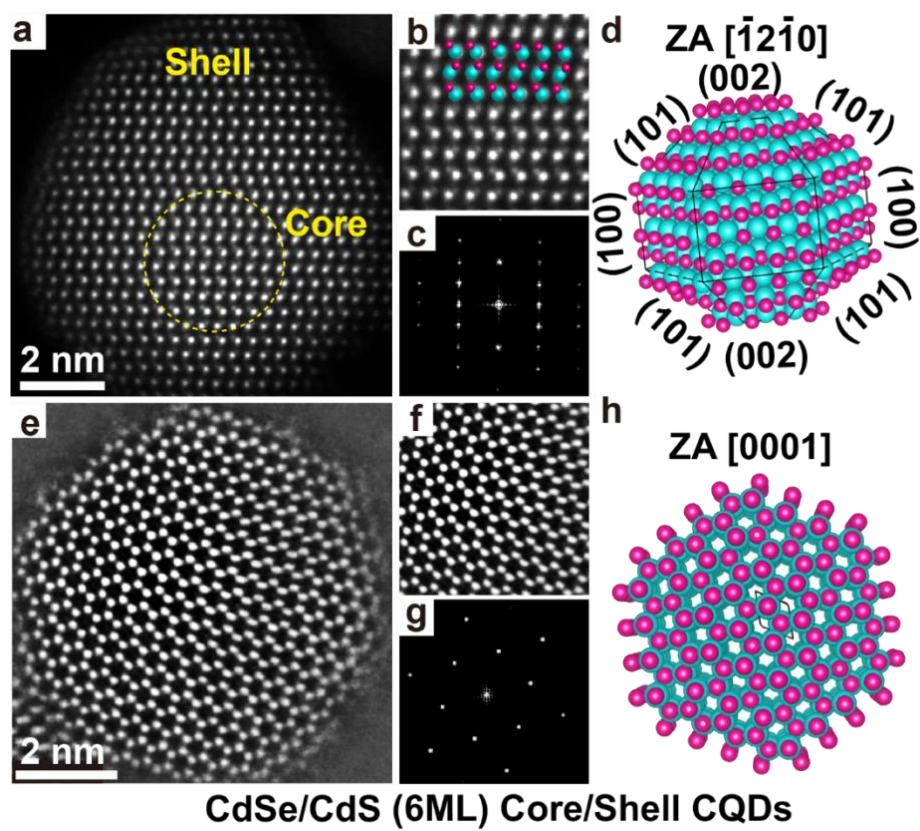

**Figure S1. Structural Characterization of the WZ-CQD monomers**. Raw (a) HAADF-STEM images of wurtzite CdSe/CdS core-shell QDs viewed under zone axis (ZA) [$\bar{1}2\bar{1}0$] (top). The magnification for the core location and fast Fourier transform (FFT) pattern analysis are shown in (b-c), respectively. The atomic structure model (d) with major low index atomic planes (002), (101), (100) under ZA [$\bar{1}2\bar{1}0$] are clearly observed thus providing identification of boundary faces of a monomer crystal. Raw (e) HAADF-STEM images of wurtzite CdSe/CdS core-shell QDs viewed under ZA [0001] (down). The magnification and FFT analysis are shown in (f-g). The atomic structure model under ZA [0001] are depicted in (h).

The atomic structure model of the QDs core-shell building blocks was identified base on the HAADF-STEM images and FFT pattern analysis. As depicted in Figure S1a-d, a typical wurtzite CdSe/CdS with low index planes (002), (101), and (100) are clearly viewed under ZA [$\bar{1}2\bar{1}0$], yielding the related atomic structure model (Figure S1d). Additionally, the corresponding raw HAADF-STEM images under ZA [0001] were shown in Figure S1e, that is, a perfect hexagonal motif was observed, which was matched well with the simulated structure model. Hence, the wurstzite CQDs monomer atomic structure model was identified for the further understanding of the binding relationship during the fusion step.

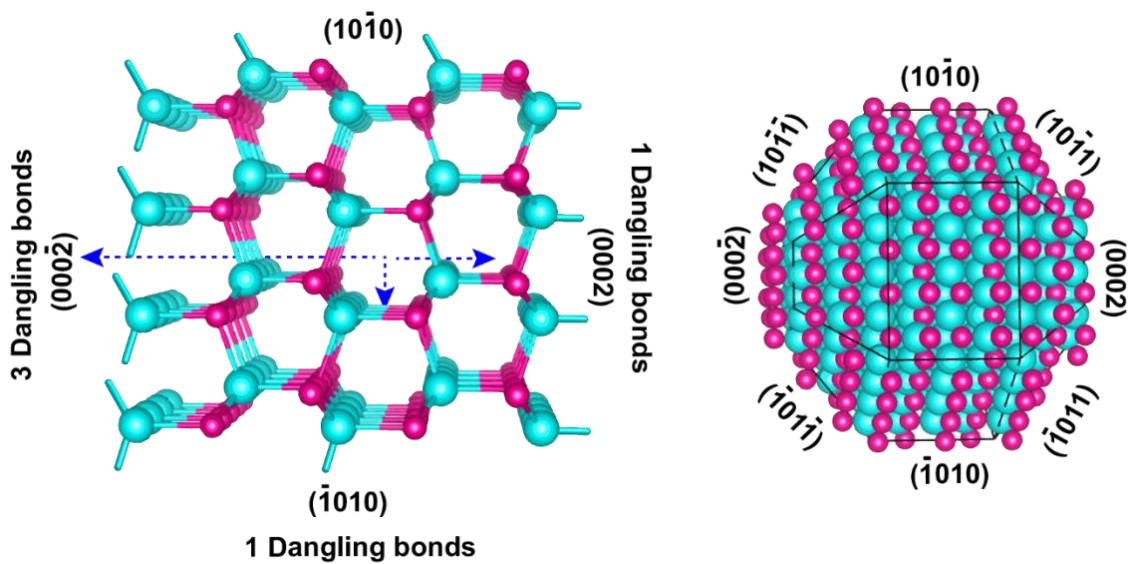

**Figure S2. Surface facets of wurtzite structure.** (000$\bar{2}$) facet: terminal Cd atoms have 3 dangling bonds, resulting in highly reactive, faster growth, and faster ligand exchange. Side facet (10$\bar{1}$0), (10$\bar{1}$1): terminal Cd atoms have 1 dangling bond with similar concentration of Cd and S sites, yielding slowest growth, slowest ligand exchange. (0002) facet: terminal Cd atoms have 1 dangling bond with higher stability, slower growth, and slower ligand exchange.

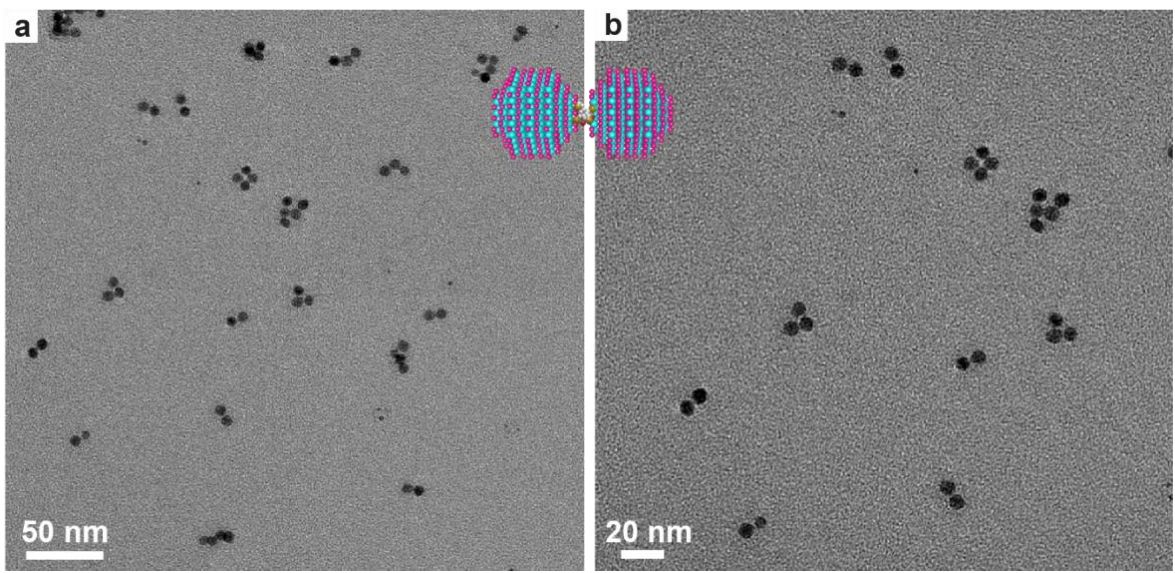

**Figure S3. Structural characterization for non-fused CQD dimers.** Low (a) and high (b) magnification of TEM images for the cross-linked CQD (non-fused) dimers.

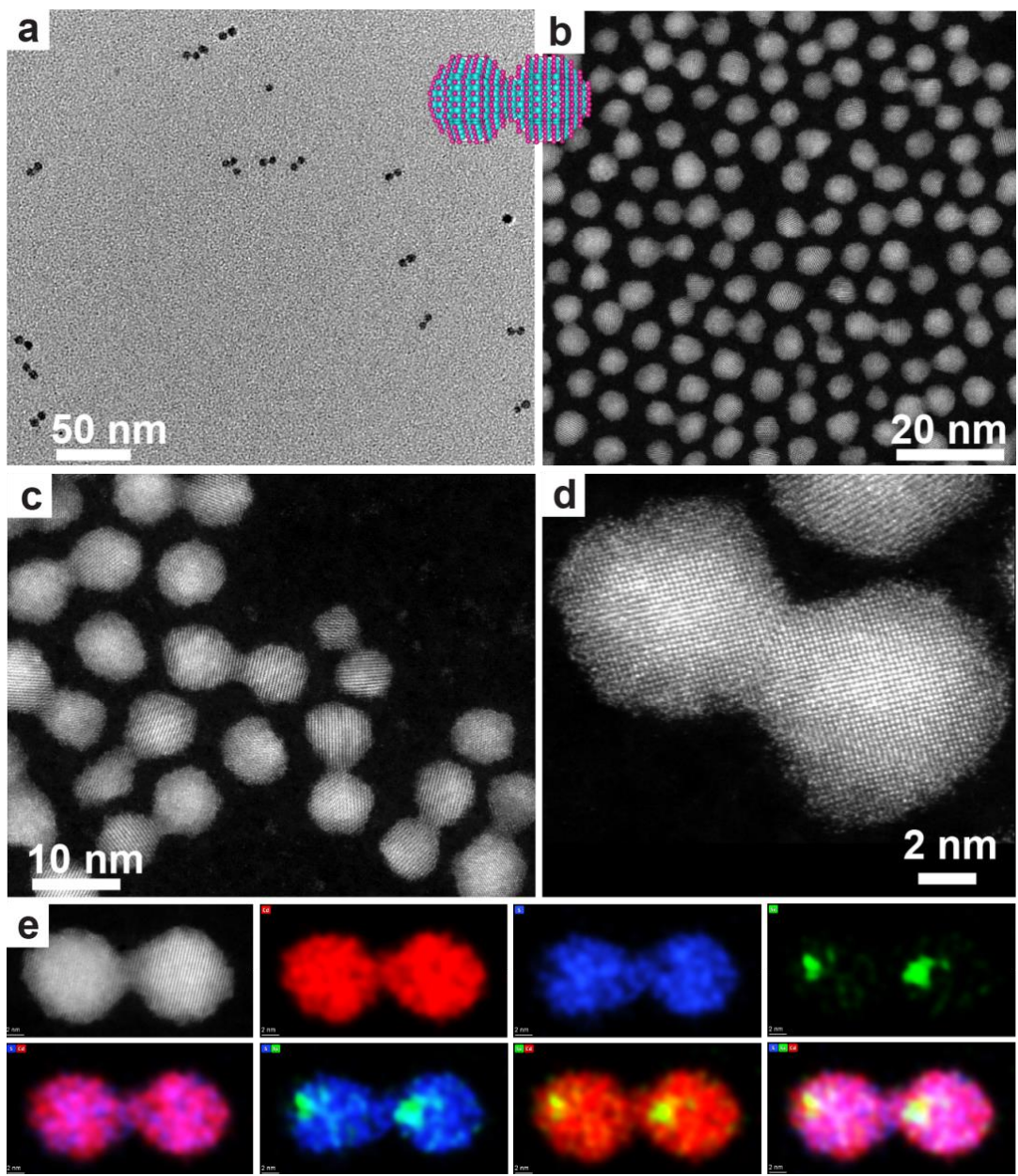

**Figure S4. Structural characterization for weakly fused CQD dimers.** TEM images of weakly fused CQD CdSe/CdS dimers (a). HAADF-STEM images with different magnification (b-e) and HAADF-STEM-EDS analysis of the weakly fused CQDs dimer with average neck diameter ~3nm.

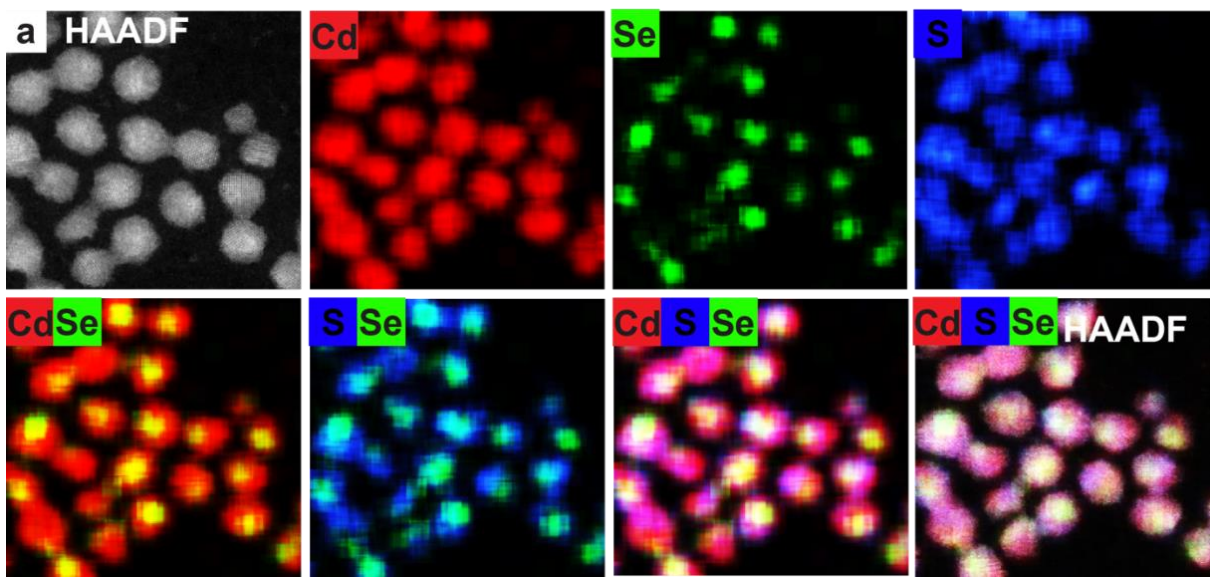

**Figure S5. STEM-EDS analysis of the weakly fused CQD dimers.** Large scale STEM-EDS elemental mapping of the weakly fused CQD dimers, showing the uniformity of the core-location distribution.

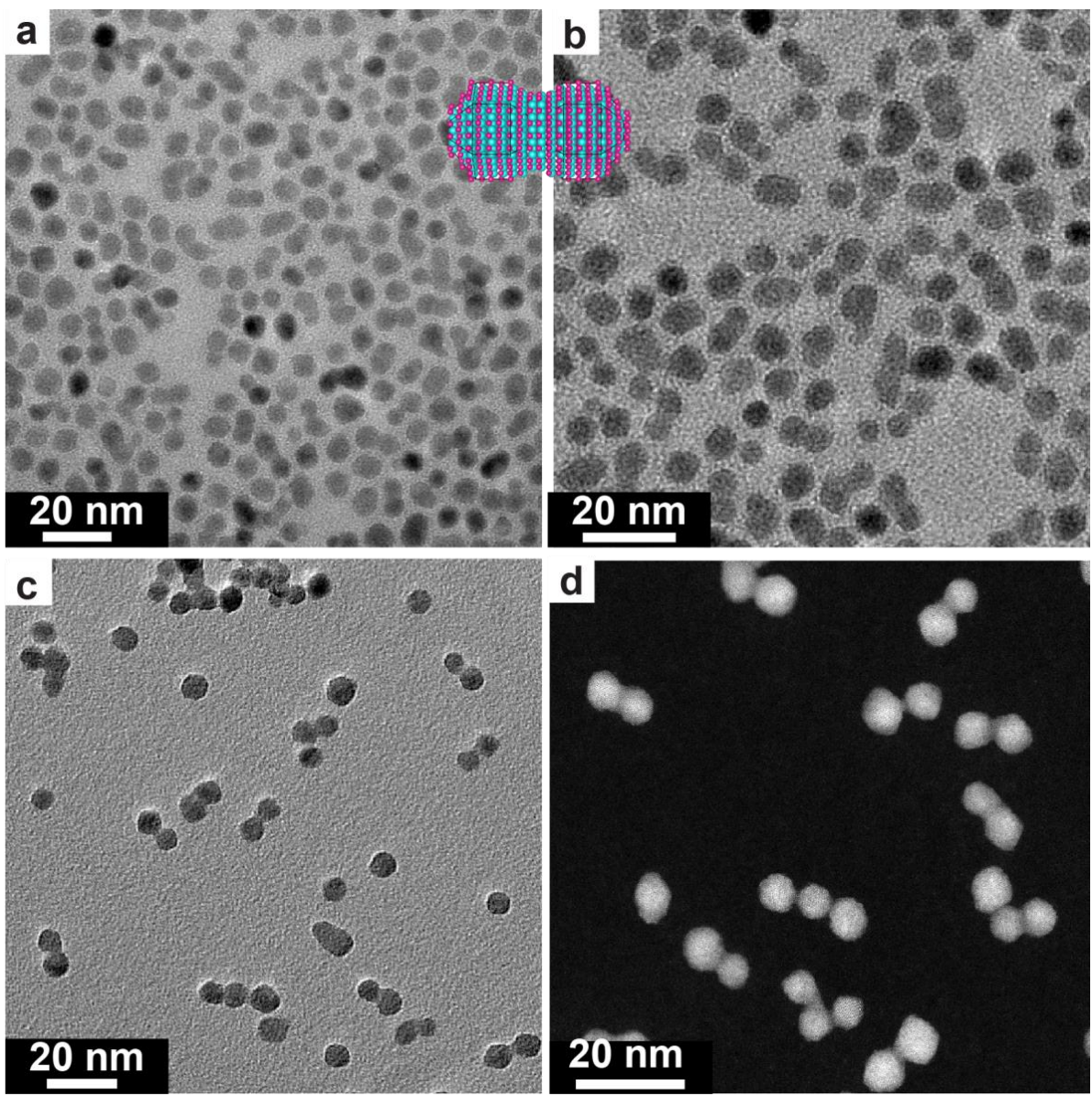

**Figure S6. Structural characterization for strongly fused CQD dimers.** TEM images with low (a), high magnification (b), HRTEM images (c), and HAADF-STEM images (d) for strongly fused CQD dimers with neck diameter ~4-4.5nm.

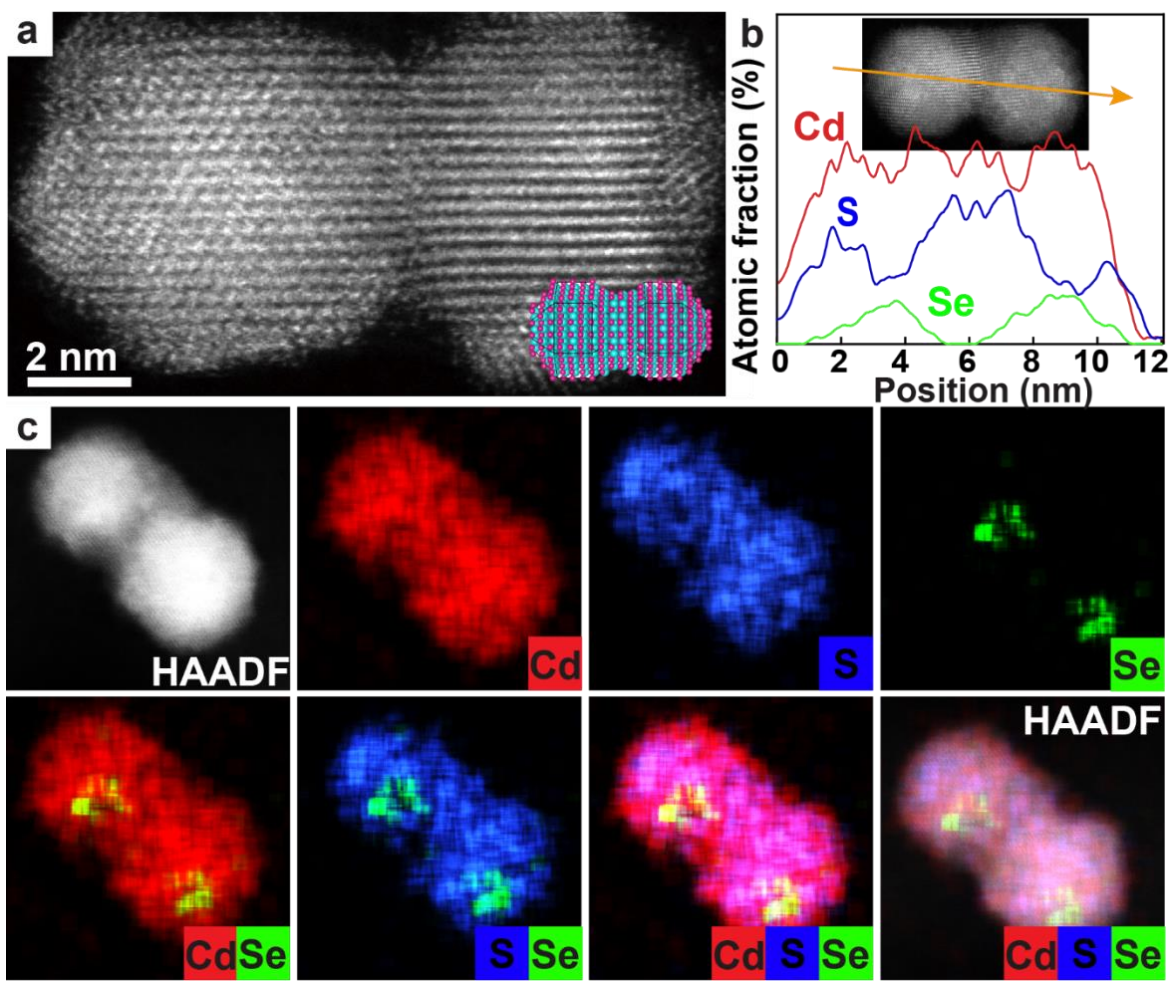

**Figure S7. Characterization for strong fused CQD dimers.** HAADF-STEM image (a), EDS line scan data (b), and STEM-EDS elemental mapping for strongly fused CQD dimers (neck diameter ~4-4.5nm). Here, the unaltered core location, and neck filling by CdS shell were clearly observed.

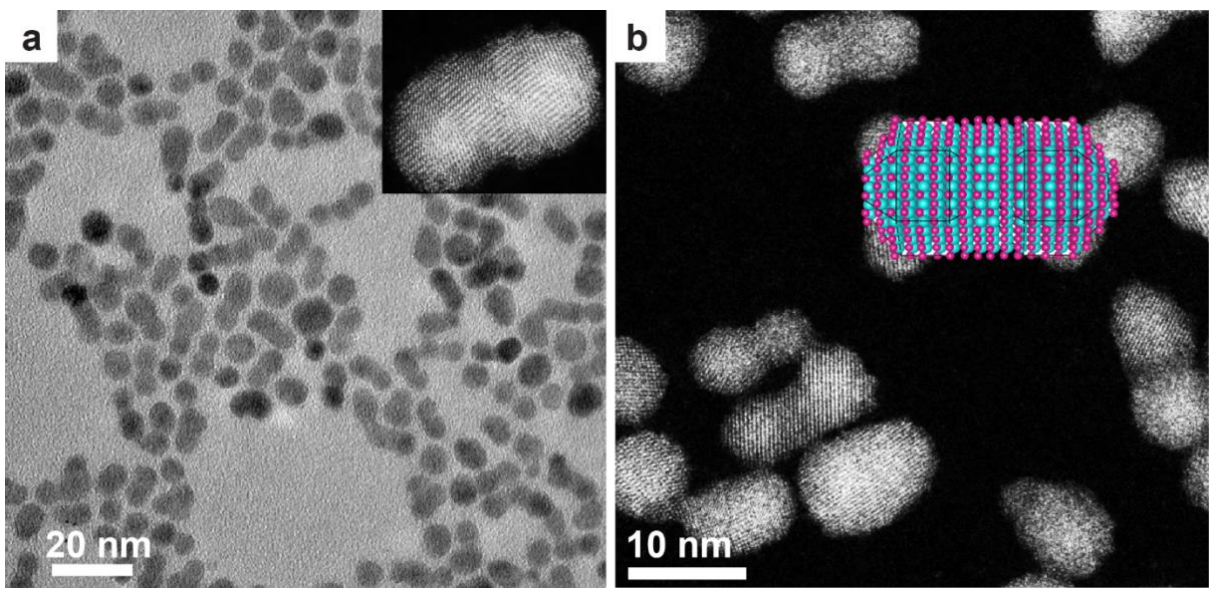

**Figure S8. Characterization for rod-like CQD dimers.** (a) Large scale HRTEM image and (b) HAADF-STEM image for rodlike CQD dimers (Neck diameter ~6-6.5nm).

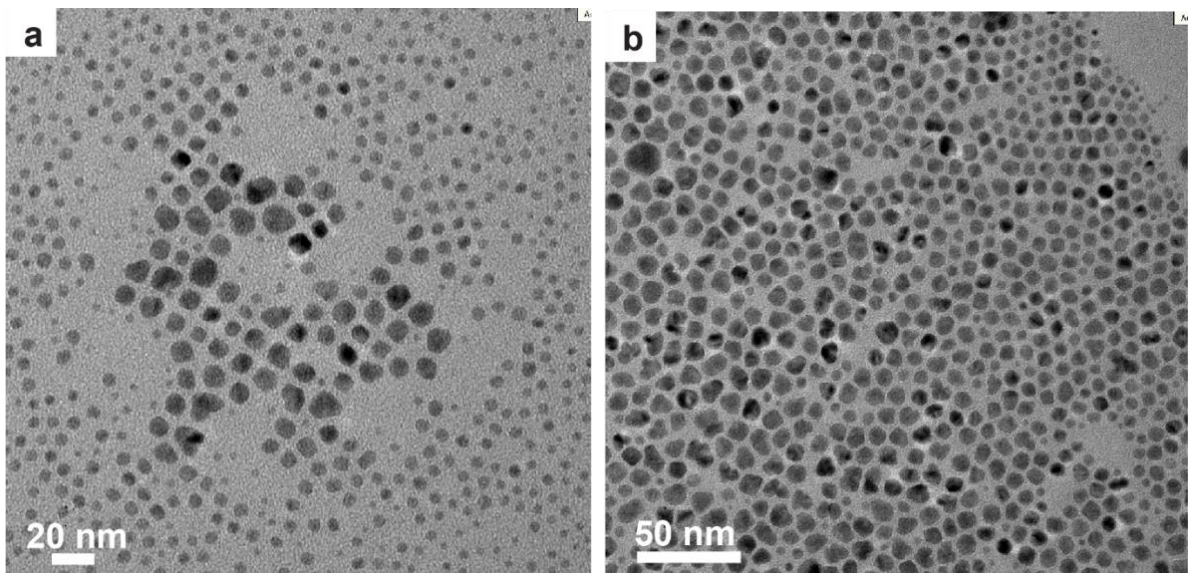

**Figure S9. Inter-particle ripenning of the CQD dimer molecules.** TEM images with different magnification (a-b) for the fused dimer with inter-particle ripenning when the temperature more than 240 °C. Here the fusion temperature is 260 °C.

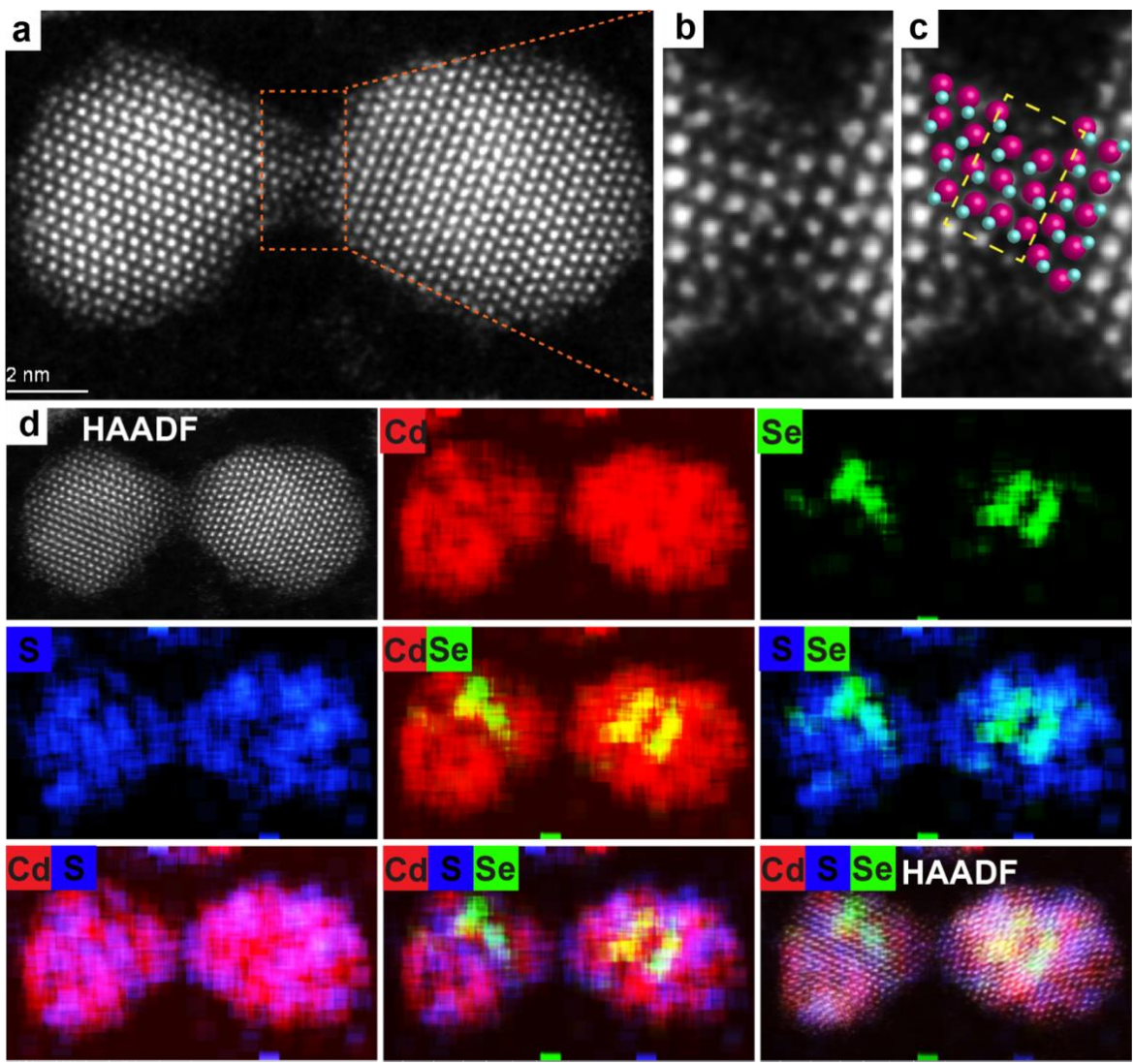

**Figure S10. Heteronymously attached CQD dimer molecule undergoing robust fusion condition.** HAADF-STEM (a-c) image and EDS elemental mapping (d) shows the neck of this types of dimers are not filled. Although the core location and the shell properties remain same as weakly fused dimer set.

The fusion of nanocrystals based on the oriented attachment (OA) mechanism to form novel structures has been widely reported. During OA in free solution, two adjacent nanoparticles could be integrated into a single nanocrystal based on the formation of the atomically matched bond between the two specific facets. Importantly, the OA procedure usually occurs between the most reactive facets to reduce the surface energy. However, the formation of the CQDs molecules in our system is based the constrained attachment, that is, the initial attachment is determined by the facets bound with the cross-linker, which shows significant difference with the typical OA. Owing to the limited rotational freedom during the two different facets attachment, stacking fault and defects may be observed in the interface. We can clearly observe the strain and dislocation of the interfacial atoms compared to the atoms arranged within both of the nanocrystals. The nanocrystals with heteronymous planes attach with an imperfect and discontinuous interface to release the strain energy. As Depicted in the Figure S10, the stacking fault and defect are observed in the high magnification HAADF-STEM imaging (Figure S10b) with atomic structure model (Figure S10c).

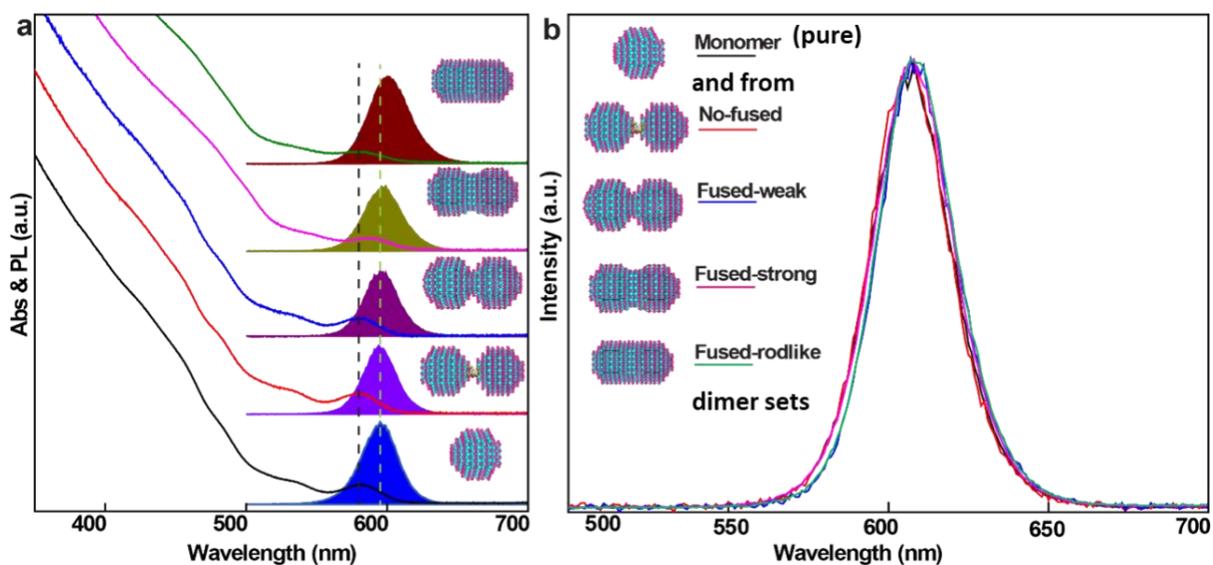

**Figure S11. Spectral characterization of different types of fused CQDs dimer.** Comparison between the spectral properties of each types of dimer sets (a) and the separated monomers from the corresponding dimer set (b). While the dimer exhibited systematic red-shift in the emission spectrum with neck filling, the separated monomer from all the sets exhibit similar spectral position and width.

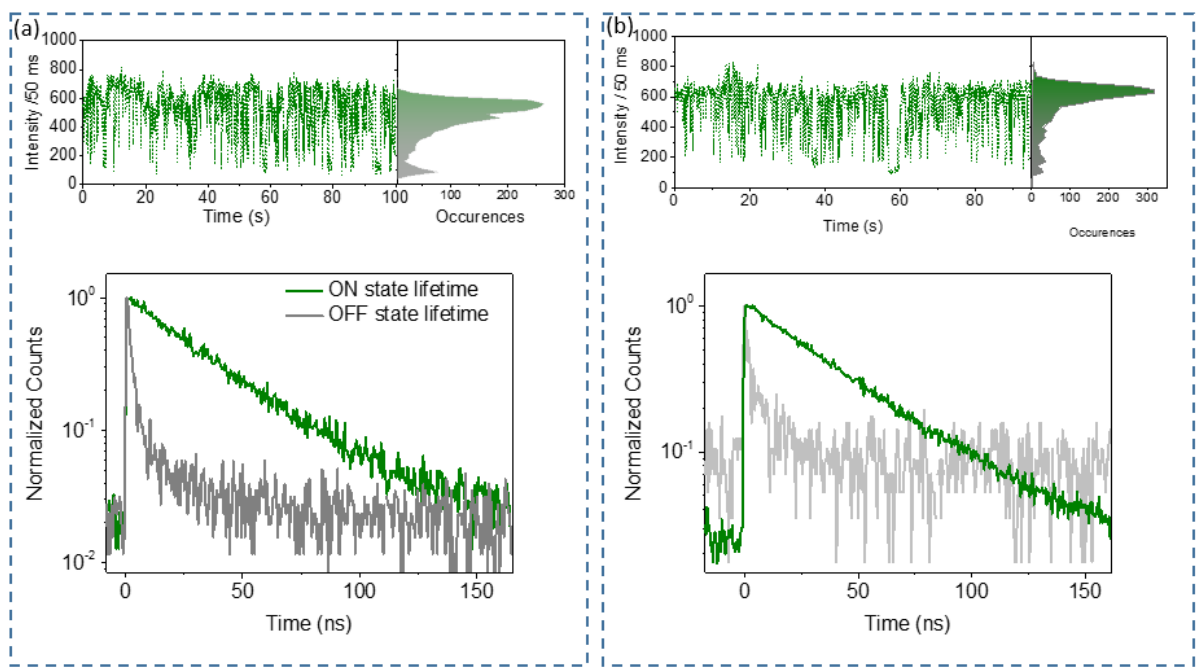

Figure S12. **Evidence of unaltered monomer properties after intra-particle ripening**. Comparison of Single Particle Characteristics of (a) pure monomer, and (b) monomer separated from rod-like dimers. Both of the monomer samples exhibit conventional bimodal distribution of emission due to ON-OFF blinking, long single exponential lifetime of the ON state, short non-radiative lifetime of the OFF states.